\documentclass[11pt]{article}

\usepackage[final]{acl}
\usepackage{times}
\usepackage{latexsym}

\usepackage[T1]{fontenc}
\usepackage[utf8]{inputenc}

\usepackage{microtype}

\usepackage{inconsolata}

\usepackage{graphicx}

\usepackage{enumitem}
\setlist{nosep}

\usepackage{booktabs}
\usepackage{multirow}
\usepackage{subcaption}
\usepackage[skip=2pt]{caption}

\usepackage{xcolor}
\usepackage{listings}

\usepackage{microtype}

\usepackage{titlesec}

\titlespacing*{\section}{0pt}{1.2ex plus .2ex minus .2ex}{0.8ex}
\titlespacing*{\subsection}{0pt}{1.0ex plus .2ex minus .2ex}{0.6ex}
\titlespacing*{\subsubsection}{0pt}{0.8ex plus .2ex minus .2ex}{0.4ex}

\definecolor{jsonbg}{RGB}{245,245,244}
\definecolor{jsonkey}{RGB}{0,92,175}
\definecolor{jsonstring}{RGB}{163,21,21}
\definecolor{jsonnumber}{RGB}{128,0,128}
\definecolor{jsonpunct}{RGB}{100,100,100}
\definecolor{jsondelim}{RGB}{20,20,20}

\lstdefinelanguage{json}{
    basicstyle=\ttfamily\small,
    showstringspaces=false,
    breaklines=true,
    frame=single,
    backgroundcolor=\color{jsonbg},
    string=[s]{"}{"},
    stringstyle=\color{jsonstring},
    literate=
     *{:}{{{\color{jsonpunct}{:}}}}{1}
      {,}{{{\color{jsonpunct}{,}}}}{1}
      {\{}{{{\color{jsondelim}{\{}}}}{1}
      {\}}{{{\color{jsondelim}{\}}}}}{1}
      {[}{{{\color{jsondelim}{[}}}}{1}
      {]}{{{\color{jsondelim}{]}}}}{1},
}

\usepackage{placeins}

\title{DUPLEX: A Dual-Plane Architecture for Monitored and Staged Repo-Level Code Repair}

\author{
\textbf{Yinuo Deng}\thanks{These authors contributed equally to this work.}\\
\normalfont UNSW Sydney, Australia\\
\normalfont\texttt{yinuo.deng1@unsw.edu.au}
\And
\textbf{Kaiyu Deng}\footnotemark[1]\\
\normalfont University of Technology Sydney, Australia\\
\normalfont\texttt{kaiyu.deng@student.uts.edu.au}
\AND
\textbf{Hanchen Wang}\\
\normalfont University of Technology Sydney, Australia\\
\normalfont\texttt{hanchen.wang@uts.edu.au}
\And
\textbf{Longbin Lai}\\
\normalfont Alibaba Group, China\\
\normalfont\texttt{longbin.lailb@alibaba-inc.com}
\AND
\textbf{Shiqi Yang}\\
\normalfont Independent Researcher\\
\normalfont\texttt{17shiqiyang@gmail.com}
\And
\textbf{Ying Zhang}\\
\normalfont University of Technology Sydney, Australia\\
\normalfont\texttt{ying.zhang@uts.edu.au}
}

\begin{document}
\maketitle
\begin{abstract}
Recent advances in large language model (LLM)-driven agents have significantly improved software repair automation. While state-of-the-art LLMs achieve strong results, mainstream LLMs are more affordable but often struggle with repo-level repair due to limited controllability and weak long-horizon progress management. We introduce \textsc{DUPLEX}, an end-to-end repo-level repair agent with a two-plane architecture: a Monitor Plane for stagnation detection and context backoff, and a Worker Plane for staged repair under independent contexts with controlled communication. On SWE-bench Verified, \textsc{DUPLEX} improves repair accuracy by 10.0\% over baseline methods with open-source models. These results demonstrate that agent architecture can substantially improve repo-level repair performance across open-source models of different scales.
\end{abstract}

\section{Introduction}
Software repair is increasingly critical as codebases grow in scale and complexity. Typically, a bug is reported as an issue with a description. Upon reproducing, developers fix it through one or more commits and submit a pull request \cite{liu2019automatic, bludau2022pr}. Manual debugging and repair are time-consuming \cite{hirsch2020fault, li2026grounded}, and bug resolution is often fragmented and inefficient due to ambiguous natural language communication \cite{al2025llput} and limited human resources.

Recent advances in large language models (LLMs) \cite{achiam2023gpt, guo2025deepseek} have enabled automated software repair by interpreting bug descriptions and generating fixes \cite{xia2024automated}. While LLMs can reduce manual effort, they remain limited on complex, multi-step tasks that require environment interaction. Agentic coding uses LLMs within a structured framework to autonomously manage software repair, including code generation, context management, task execution, and decision-making \cite{yang2024swe, zhang2024autocoderover, ma2024understand}.

Agentic coding allows an LLM to work on software engineering tasks similarly to a human \cite{jimenez2023swe}. 
The LLM generates code and instructions, while the agent executes them and returns feedback in an agentic loop \cite{yao2022react}. 
Despite their promise, existing agentic systems often suffer from coarse-grained context management and overreliance on the underlying LLM for decision-making.  
% decision-making that is highly dependent on the intrinsic capabilities of the underlying LLM. 
In practice, context is often propagated across stages without relevance-aware selection, weakening its effectiveness, and the agent’s trajectory is largely shaped by the LLM’s step-by-step reasoning, making the system sensitive to model limitations and randomness. 

Repo-level tasks present more challenges than generating code snippets \cite{xia2025demystifying}. We identify the following practical challenges: \textbf{(1) Focus loss.} As the agent proceeds through a complex repair task, it may lose track of earlier requirements \cite{liu2024lost, wang2025breaking}. 
Different stages also have different objectives, shifting the agent’s focus. 
\textbf{(2) Progress stagnation.} When an operation fails, the agent may repeat unproductive attempts, adding irrelevant context that hinders progress on the core task. 
\textbf{(3) Context relevance imbalance.} Not all context is equally important. High-level strategies, such as plans, are critical, but low-level details from editing and testing are only weakly relevant and can dilute decision-making focus. \textbf{(4) Lack of post-process quality assurance.} During execution, the agent may generate temporary files, such as scripts \cite{lee2025sec}, and produce code that leaves room for structural improvements \cite{dong2025code, sun2025quality}. 
Current agents lack a post-repair process for cleanup and quality assurance, resulting in unnecessary complexity in the output \cite{wadhwa2024core}.

To address these challenges, we propose \textsc{DUPLEX}\footnote{\url{https://github.com/dynos01/duplex}.}, an end-to-end framework for repo-level repair that regulates agent behavior and improves long-horizon progress. \textsc{DUPLEX} features a two-plane architecture: a Monitor Plane and a Worker Plane. The Monitor Plane tracks repair progress, detects stagnation, and triggers context backoff when necessary. The Worker Plane has three stages: Reproduction, Implementation, and Quality Assurance, each with independent contexts to preserve stage-specific focus and limit irrelevant history carryover. It also uses a controlled communication protocol that delegates low-level actions to a lightweight sub-agent, reducing context pollution, and supports cross-stage querying and consensus. The contributions of our work are as follows:

\begin{itemize}
    \item We analyze trajectories of widely used software engineering agents to identify weaknesses in task understanding and behavior control, providing insights for improvement.
    \item We propose \textsc{DUPLEX}, an end-to-end repo-level repair agent with a two-plane architecture: a Monitor Plane for stagnation detection and context backoff, and a Worker Plane for staged repair under independent contexts with controlled communication.
    \item We evaluate \textsc{DUPLEX} on SWE-bench Verified \cite{jimenez2023swe} and demonstrate improved end-to-end performance.
\end{itemize}

\section{Trajectory Analysis}\label{s2}

\begin{figure*}[t]
    \centering
    \includegraphics[width=6in]{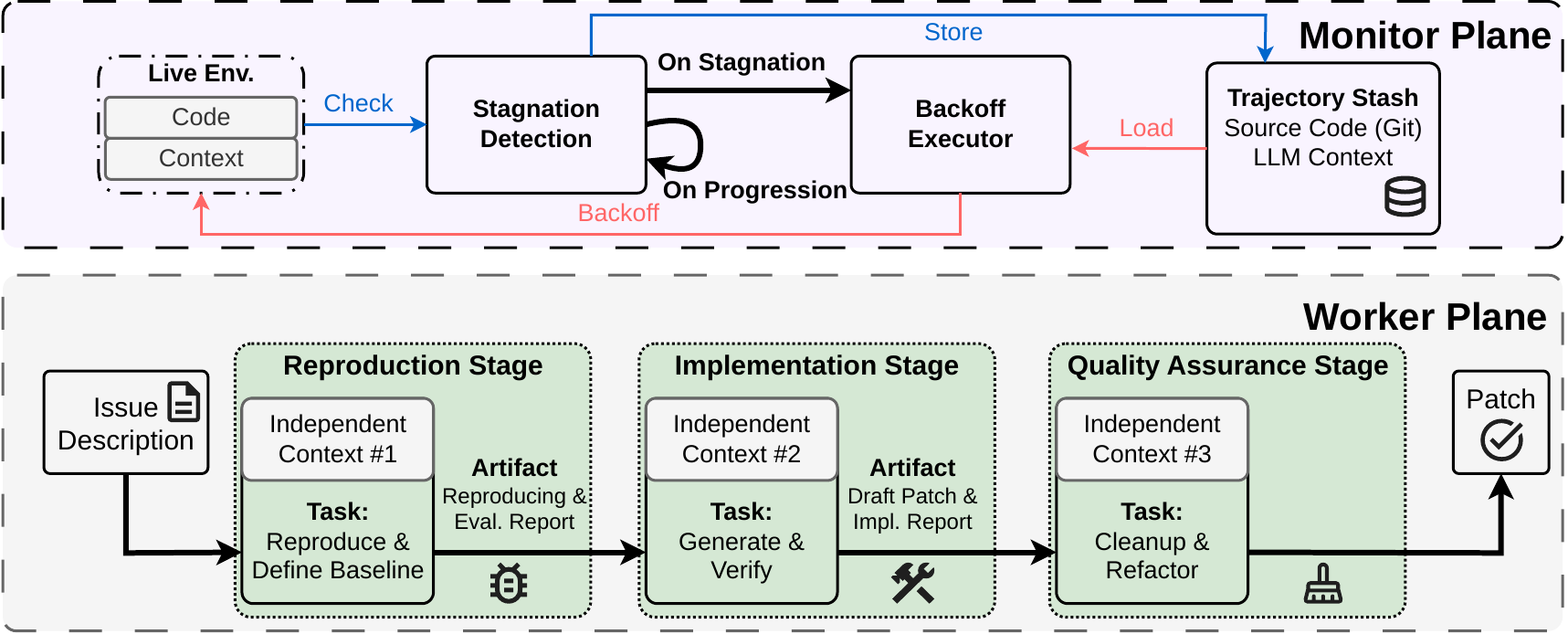} 	
    \caption{The architecture of \textsc{DUPLEX}.}
    \label{fig_overview}
\end{figure*}

\begin{lstlisting}[language=json, caption={Stagnation (\texttt{django\_\_django-11400)}}, label=exp_1]
{
    "role": "assistant",
    "content": "THOUGHT: There are still issues with the formatting. Let me try to fix the indentation and line break manually:\n\n```bash\nsed -i '482s/foo.*/bar/' some/path.py```"
} ... (repeats 99 times)
\end{lstlisting}

In this section, we analyze trajectories from the public SWE-bench evaluation records\footnote{\url{https://github.com/SWE-bench/experiments}} to identify recurring failure patterns in repo-level repair. We focus on runs produced by \texttt{mini-swe-agent} with \texttt{Qwen3-Coder-480B-A35B-Instruct} as the model. According to the published records, this run resolves 277 out of 500 instances (55.4\%). Our analysis focuses on these unresolved ones. Rather than treating failure as merely an outcome, we examine the dominant behaviors that repeatedly appear before failure. The analysis reveals three common issues: stagnation in local loops, degradation of prompt compliance over long horizons, and final patches that are not review-ready. These observations directly motivate our design.

\paragraph{Finding 1: Agent trapped in long local loops (Example~\ref{exp_1}).} Among the 223 unresolved instances, 211 (94.6\%) exhibit clear repeated local behavior, with 205 (91.9\%) showing particularly persistent patterns of repetition across multiple steps. These loops typically include repeated edits to the same file or code region, repeated execution of similar commands, repeated local testing, or prolonged fixation on a narrow part of the repository. Such behavior indicates that the agent can remain trapped in a local area for a long time without generating new evidence or making substantial progress toward a validated fix.
% \FloatBarrier

\begin{lstlisting}[language=json, caption={Rule Violation (\texttt{django\_\_django-11276)}}, label=exp_2]
{
    "content": "... DO NOT MODIFY: Tests, configuration files (pyproject.toml, setup.cfg, etc.) ..."
},
{
    "content": "THOUGHT: I need to update line 30 in the test file to expect `&#x27` instead of `&#39`. Let me make that change:\n\n```bash\nsed -i '30s/&#39;/\\&#x27;/' ./tests/utils_tests/test_html.py\n```"
}
\end{lstlisting}

\paragraph{Finding 2: Prompt compliance degrades (Example~\ref{exp_2}).} We identify 113 failed instances (50.7\% of unresolved cases) whose first explicit rule violation occurs near the beginning. 61 cases produce multiple bash blocks despite a one-block requirement, 19 omit the required bash block, 1 omits the required \texttt{THOUGHT} field, and 32 modify repository test files despite explicit instructions not to do so. This pattern suggests that prompt-following failures are not only caused by poor initial understanding, but also by gradual control loss as the context grows and high-level constraints become diluted by accumulated interaction history.

\begin{lstlisting}[language=json, caption={Bad Patch (\texttt{sphinx-doc\_\_sphinx-9229)}}, label=exp_3]
{
    "submission": "diff --git a/_build/html/.doctrees/CONTRIBUTING.doctree b/_build/html/.doctrees/CONTRIBUTING.doctree\nnew file mode 100644\nindex 000000000..089ce55f7\nBinary files /dev/null and b/_build/html/.doctrees/CONTRIBUTING.doctree differ..."
}
(1,136 changes, 367 binaries, len = 15,271,127)
\end{lstlisting}

\paragraph{Finding 3: Patches are often not review-ready (Example~\ref{exp_3}).} Among the 223 unresolved cases, 202 (90.6\%) have a patch, and 162 of these patches (80.2\%) contain non-essential artifacts. Common examples include temporary debugging scripts, generated files, and other auxiliary outputs that should not appear in a final submission. This pattern indicates that producing a candidate fix is not sufficient on its own: a practical repair agent also needs a dedicated post-repair process for cleanup, consolidation, and patch-quality improvement.

\paragraph{Implication for our design.} These findings suggest that a practical repo-level repair agent needs explicit trajectory monitoring to detect stagnation, context management that prevents long, noisy histories from overwhelming the model, and a dedicated post-repair stage that cleans temporary artifacts and improves patch quality. These observations directly motivate the Monitor Plane and the three-stage Worker Plane of \textsc{DUPLEX}.

\section{The \textsc{DUPLEX} Architecture}\label{s3}

In this section, we present \textsc{DUPLEX} to address the challenges identified in Section \ref{s2}.
Figure \ref{fig_overview} shows the architecture of \textsc{DUPLEX}, which consists of two logical planes: a Monitor Plane and a Worker Plane. The two planes interact in an interleaved manner: after each LLM invocation in the Worker Plane, the Monitor Plane inspects the current context and may edit it before the next step.

\subsection{Design Rationale}
\textsc{DUPLEX} is motivated by real-world software development workflows involving both supervision and execution. Developers typically first reproduce and understand the issue, then implement a fix, and finally validate the patch for integration \cite{vyas2014bug, li2026grounded}.
Effective repair also depends on a higher-level process that monitors progress and decides when to revise direction. These responsibilities are distinct and require different forms of reasoning.

This motivates our two-plane architecture. The Worker Plane handles execution and is divided into three stages (Reproduction, Implementation, and Quality Assurance) that reflect the natural phases of debugging and patch development. Each stage operates in an independent context to maintain focus and reduce interference from irrelevant details accumulated elsewhere. In contrast, the Monitor Plane oversees the long-horizon repair trajectory, detects stagnation, and triggers context backoff when necessary. Inspired by collaborative engineering practice, \textsc{DUPLEX} allows stages to delegate low-level editing and verification to a lightweight sub-agent, and enables later stages to query or challenge key conclusions from earlier stages before advancing. Together, these choices align the agent with real software repair workflows and improve focus, controllability, and robustness in repo-level tasks.

\subsection{Monitor Plane}\label{s3.2}
The Monitor Plane provides trajectory-level control of the repair process. 
Its main purpose is to detect \textit{stagnation}, where the agent stays stuck in a local problem area for too long and repeatedly makes attempts without substantial progress.
Here, stagnation does not simply mean the bug remains unfixed; rather, it refers to a recent trajectory that fails to generate new evidence, improve understanding of the fault, or move the repair process closer to a validated patch. 
This behavior is common in repo-level repair, where repeated edits, command executions, and verification attempts can quickly accumulate as noise and weaken later decision-making.

To address this issue, the Monitor Plane periodically evaluates recent progress while remaining interleaved with the Worker Plane. Specifically, it is invoked every $m$ Worker Plane steps. For periodic progress assessment, the Monitor Plane uses two parameters: an inspection interval $m$ (default $m=5$), which specifies how often to check progress, and a lookback window $n$ (default $n=15$), which specifies how many recent steps to consider. At each inspection point, the Monitor Plane evaluates the most recent $n$ steps through a dedicated evaluation LLM call to determine whether they contain substantial progress. When stagnation is detected, the Monitor Plane triggers a \textit{backoff} operation. The evaluator defaults to keeping the current trajectory and considers rollback only for strict repeated error loops, sustained task drift, or unrecovered source corruption. Ordinary command failures or strategy changes alone do not trigger rollback. When rollback is selected, it chooses a valid step-aligned checkpoint and provides recovery advice, while invalid targets or malformed judgments default to keeping the current state.

A backoff operation restores the agent to an earlier checkpoint containing both the \textit{context state} and the \textit{source-code state}. The context state is recovered by truncating the interaction history to the target step, removing recent low-value reasoning and execution noise. The source-code state is recovered using \texttt{git}: each agent step produces a commit, and the Monitor Plane tracks the mapping from trajectory steps to commits. When backoff is triggered, the repository is reset to the selected step using \texttt{git}, while the context is rolled back to the same point. By jointly reverting context and code, the Monitor Plane lets the agent resume exploration from a clean state and prevents ineffective local attempts from dominating future decisions.

\subsection{Worker Plane}
The Worker Plane handles the execution side of repo-level repair. It organizes repair into three stages: Reproduction, Implementation, and Quality Assurance. These stages have distinct objectives and therefore use dedicated contexts rather than a shared one. Compared with monolithic agent designs, this decomposition reduces interference from irrelevant history and provides a clear scheme for artifact handoff and controlled communication. 

Each stage in the Worker Plane uses an independent context rather than inheriting the full history from previous stages. This design reflects that the three stages pursue different objectives and therefore require different working memories. A context useful for constructing a reproduction script may be far less relevant when implementing the fix, while implementation-specific reasoning may be unnecessary or even distracting during quality assurance. Therefore, when a new stage begins, its context is reset and initialized with a stage-specific prompt to re-anchor the agent to the current objective.

To transfer knowledge across stages without carrying over the full context, \textsc{DUPLEX} adopts a structured handoff mechanism. At the end of each stage, the agent stores outputs such as scripts and reports in designated locations. These condensed artifacts serve as task-relevant knowledge that allows the next stage to quickly recover the problem and begin substantive work in a clean context.

The three stages are as follows:

\begin{itemize}[leftmargin=*, itemsep=2pt, topsep=2pt]
    \item \textbf{Reproduction Stage} bridges the gap between the natural-language issue description and a machine-verifiable specification. After interpreting the issue and exploring the repository, it produces a test script as the canonical reproduction baseline, along with a report that reformulates the issue in a more technical and operational manner.
    \item \textbf{Implementation Stage} produces a fix that satisfies the reproduction baseline. Before modifying the codebase, it reviews the test script and reproduction report to confirm the repair direction, and may request clarification or locally refine its working interpretation of these artifacts if further auditing is needed. It concludes once the candidate patch passes the reproduction script.
    \item \textbf{Quality Assurance Stage} refines the candidate fix into a high-quality final patch. The repair process may leave behind temporary scripts, unused code, auxiliary files, or other artifacts that should not appear in the final submission. Moreover, LLM-generated code may be functionally correct but structurally non-idiomatic. This stage performs cleanup and refactoring, including removing unnecessary files such as binaries.
\end{itemize}

While this decomposition improves focus, a purely static workflow would oversimplify real workloads. Practically, later phases may need to revisit earlier interpretations, and high-level reasoning requires low-level execution support. A strict pipeline risks losing flexibility. Therefore, \textsc{DUPLEX} incorporates a communication mechanism in the Worker Plane, which serves two purposes: \textbf{controlled later-to-earlier stage consultation} when clarification or reconciliation is needed, and \textbf{low-level operation delegation} to a lightweight sub-agent so that the main stage agent remains focused on high-level reasoning rather than tool-specific parameters.

\begin{figure}[ht]
    \centering
    \includegraphics[width=\columnwidth]{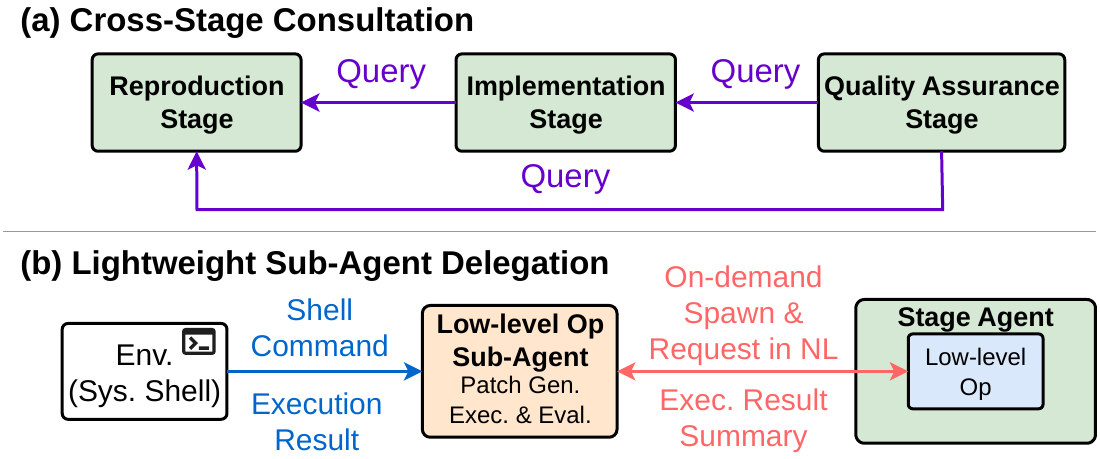}
    \vspace{2mm}
    \caption{Communication mechanism of \textsc{DUPLEX}.}
    \label{fig:comm}
\end{figure}

\paragraph{Cross-stage consultation.} Although stages in \textsc{DUPLEX} run in separate contexts, later stages may still need to query or challenge earlier conclusions. For example, later stages may question the issue interpretation, and therefore need a way to query or challenge the previous stages before proceeding. To support this, we retain a completed stage as a \textit{consultable reasoning unit} after execution ends. When such a query arises, the system provides the LLM with the preserved context, a question-answering prompt, and the query. From the agent's perspective, the previous stages remain available only in a query-only role: they no longer interact with the environment or execute commands, and only answer questions about their earlier reasoning. This process is shown in the upper part of Figure \ref{fig:comm}. This design preserves context isolation across stages while allowing later stages to recover missing rationale and reconcile interpretations before proceeding.

\paragraph{Lightweight sub-agent delegation.} In conventional ReAct-style agents, code editing, command execution, and result verification are typically performed in the same context as exploration and reasoning. However, in repo-level repair, these low-level operations are often only instrumental to a higher-level goal, such as testing whether a fix works. Moreover, because LLM-generated editing instructions are not always precise on the first attempt, editing and verification may require multiple retries, causing low-value operational details to accumulate in the main context. To reduce this interference, \textsc{DUPLEX} introduces a lightweight sub-agent for source-code editing and validation.

As shown in the lower part of Figure \ref{fig:comm}, when a stage needs to modify a file and check the outcome, the main stage agent communicates the intended edit and the validation procedure to the sub-agent in natural language. The sub-agent then carries out low-level operations, such as applying edits, running commands, and collecting outputs, and reports the result to the main stage agent. The parent agent may accept the result, discard it, or request rework, while the sub-agent may also ask the parent stage for clarification. When a proposed modification is accepted, the parent stage records a brief state update before continuing, summarizing the current project status, the immediate goal, and whether the current repair direction may be drifting. The sub-agent is invoked on demand and terminated once the result has been resolved. This design separates high-level reasoning from low-level action realization, allowing the main stage agent to remain focused on diagnosis, decision-making, and repair strategy while reducing context pollution from repeated operational attempts.

\begin{table}[t]
\centering
\footnotesize
\caption{Main stage agent state machine.}
\label{tab:stage_fsm}
\begin{tabular}{p{1.1cm} p{2.8cm} p{2.6cm}}
\toprule
\textbf{State} & \textbf{Allowed actions} & \textbf{Transition} \\
\midrule
Normal
& shell; consult earlier stage; spawn child; submit
& spawn $\rightarrow$ Waiting; submit $\rightarrow$ Submitted \\
\midrule
Waiting
& review child report; reply queries
& accept $\rightarrow$ Reflect; discard/rework $\rightarrow$ Waiting \\
\midrule
Reflect
& state update
& done $\rightarrow$ Normal \\
\midrule
Submitted
& reply queries only
& no further exec. \\
\bottomrule
\end{tabular}
\end{table}

\begin{table}[t]
\centering
\footnotesize
\caption{Lightweight sub-agent state machine.}
\label{tab:child_fsm}
\begin{tabular}{p{1.1cm} p{2.8cm} p{2.6cm}}
\toprule
\textbf{State} & \textbf{Allowed actions} & \textbf{Transition} \\
\midrule
Active
& shell; ask/challenge parent; report
& report $\rightarrow$ Reported \\
\midrule
Reported
& await parent decision
& accept/discard $\rightarrow$ Terminated; rework $\rightarrow$ Active \\
\midrule
Terminated
& none
& end \\
\bottomrule
\end{tabular}
\end{table}

\begin{table*}[t]
\centering
\small
\caption{Per-repository evaluation results of RQ1 on SWE-bench Verified. Each cell reports the success rate. \textbf{Qwen} = \texttt{Qwen3-Coder-480B-A35B-Instruct}, \textbf{GPT} = \texttt{gpt-oss-120b}, and \textbf{GLM} = \texttt{GLM-4.7-Flash}. MSA = \texttt{mini-swe-agent}. SWA = \texttt{SWE-agent}. ACR = \texttt{AutoCodeRover}. OH = \texttt{OpenHands}. OURS = \texttt{\textsc{DUPLEX}}. }
\label{tab:rq1_repo_breakdown}
\setlength{\tabcolsep}{1.1pt}
\renewcommand{\arraystretch}{1}

\begin{tabular}{llccccccccccccc}
\toprule
\textbf{Model} & \textbf{Agent} 
& astropy & django & matplotlib & mwaskom & pallets & psf & pydata & pylint & pytest & sklearn & sphinx & sympy & TOTAL \\
& 
& (22) & (231) & (34) & (2) & (1) & (8) & (22) & (10) & (19) & (32) & (44) & (75) & (500) \\
\midrule
\multirow{5}{*}{\textbf{Qwen}} & MSA & \textbf{45.5\%} & 58.0\% & \textbf{58.8\%} & \textbf{0.0\%} & \textbf{100.0\%} & 50.0\% & 68.2\% & 10.0\% & 47.4\% & 68.8\% & 36.4\% & 60.0\% & 55.4\% \\
 & SWA & 27.3\% & 48.5\% & 32.4\% & \textbf{0.0\%} & \textbf{100.0\%} & 12.5\% & 40.9\% & 10.0\% & 57.9\% & 68.8\% & 20.5\% & 50.7\% & 44.2\% \\
 & ACR & 22.7\% & 50.2\% & 35.3\% & \textbf{0.0\%} & \textbf{100.0\%} & 37.5\% & 50.0\% & \textbf{20.0\%} & 42.1\% & 53.1\% & 25.0\% & 46.7\% & 44.2\% \\
 & OH$^{\dagger}$ & 13.6\% & 65.8\% & 50.0\% & \textbf{0.0\%} & \textbf{100.0\%} & 0.0\% & \textbf{81.8\%} & \textbf{20.0\%} & 57.9\% & 81.2\% & 0.0\% & 42.7\% & 52.4\% \\
 & OURS & \textbf{45.5\%} & \textbf{66.7\%} & \textbf{58.8\%} & \textbf{0.0\%} & \textbf{100.0\%} & \textbf{75.0\%} & 77.3\% & \textbf{20.0\%} & \textbf{84.2\%} & \textbf{84.4\%} & \textbf{54.5\%} & \textbf{66.7\%} & \textbf{65.4\%} \\
\midrule
\multirow{4}{*}{\textbf{GPT}} & MSA & 18.2\% & 29.0\% & 29.4\% & \textbf{0.0\%} & \textbf{0.0\%} & 50.0\% & 27.3\% & 10.0\% & 42.1\% & 28.1\% & 11.4\% & 21.3\% & 26.0\% \\
 & SWA & 9.1\% & 22.1\% & 11.8\% & \textbf{0.0\%} & \textbf{0.0\%} & 25.0\% & \textbf{40.9\%} & 10.0\% & 31.6\% & 34.4\% & 4.5\% & 21.3\% & 20.8\% \\
 & ACR & 0.0\% & 22.1\% & 5.9\% & \textbf{0.0\%} & \textbf{0.0\%} & 0.0\% & 18.2\% & 0.0\% & 0.0\% & 6.2\% & 0.0\% & 5.3\% & 12.6\% \\
 & OURS & \textbf{36.4\%} & \textbf{39.0\%} & \textbf{35.3\%} & \textbf{0.0\%} & \textbf{0.0\%} & \textbf{62.5\%} & \textbf{40.9\%} & \textbf{20.0\%} & \textbf{47.4\%} & \textbf{37.5\%} & \textbf{20.5\%} & \textbf{45.3\%} & \textbf{38.0\%} \\
\midrule
\multirow{4}{*}{\textbf{GLM}} & MSA & \textbf{40.9\%} & 48.1\% & 35.3\% & \textbf{0.0\%} & \textbf{100.0\%} & 50.0\% & 54.5\% & 20.0\% & 63.2\% & 65.6\% & \textbf{43.2\%} & 37.3\% & 46.2\% \\
 & SWA & \textbf{40.9\%} & \textbf{51.5\%} & 23.5\% & \textbf{0.0\%} & 0.0\% & 50.0\% & \textbf{63.6\%} & 20.0\% & 68.4\% & \textbf{71.9\%} & 27.3\% & 46.7\% & 47.8\% \\
 & ACR & 27.3\% & 31.6\% & 29.4\% & \textbf{0.0\%} & \textbf{100.0\%} & 37.5\% & 36.4\% & \textbf{30.0\%} & 15.8\% & 59.4\% & 11.4\% & 21.3\% & 29.4\% \\
 & OURS & 36.4\% & 51.1\% & \textbf{47.1\%} & \textbf{0.0\%} & \textbf{100.0\%} & \textbf{62.5\%} & \textbf{63.6\%} & \textbf{30.0\%} & \textbf{73.7\%} & \textbf{71.9\%} & 40.9\% & \textbf{50.7\%} & \textbf{51.6\%} \\
\bottomrule
\end{tabular}

\raggedright
\footnotesize{$^{\dagger}$ OpenHands is a complex framework with sophisticated control flows. With lightweight models, we observe repeated parsing failures and very limited progress. Therefore, OpenHands is only reported with \texttt{Qwen3-Coder-480B-A35B-Instruct}.}
\end{table*}

\paragraph{State-machine enforcement.}
To make communication controlled rather than free-form, \textsc{DUPLEX} constrains both the main stage agent and the lightweight sub-agent with explicit finite-state interaction rules. Tables~\ref{tab:stage_fsm} and~\ref{tab:child_fsm} summarize the allowed actions and transitions for the two roles. This design keeps inter-stage consultation and delegated low-level execution focused, reviewable, and resistant to unnecessary context growth.

Taken together, these design choices make the Worker Plane a controlled yet flexible execution framework for repo-level repair. Stage decomposition and independent contexts preserve phase-specific focus, structured handoff artifacts transfer condensed knowledge, and controlled communication restores the flexibility missing in a strictly linear pipeline. As a result, the Worker Plane supports not only issue reproduction and fix implementation, but also the refinement of the final patch into a more maintainable form. Overall, the Worker Plane and Monitor Plane together establish two complementary loops. They realize an \textit{implementation-evaluation loop} by coupling staged repair and executable verification with trajectory-level progress control, and a \textit{correctness-maintainability loop} by coupling validated fixing with controlled refinement into review-ready patches.

\section{Evaluation}\label{s4}

In this section, we conduct extensive experiments on the dataset of SWE-bench Verified to evaluate the performance of \textsc{DUPLEX}. We aim to answer the following research questions:

\begin{list}{}{%
    \setlength{\leftmargin}{3.1em}%
    \setlength{\labelwidth}{2.5em}%
    \setlength{\labelsep}{0.6em}%
    \setlength{\itemsep}{0pt}%
    \setlength{\topsep}{0pt}%
}
    \item[\textbf{RQ1.}] How effective is \textsc{DUPLEX}?
    \item[\textbf{RQ2.}] How compact are \textsc{DUPLEX} patches?
    \item[\textbf{RQ3.}] How does each component of \textsc{DUPLEX} affect the end-to-end results?
\end{list}

\subsection{Experimental Setup}
\paragraph{Dataset.} We evaluate \textsc{DUPLEX} using the SWE-bench Verified \cite{jimenez2023swe} dataset, which is widely used across various works, thus providing a fair comparison base. This dataset is a human-validated subset of the original SWE-bench and contains 500 Python test cases across 12 repos. These test cases are generated from real-world GitHub issues and corresponding pull requests.

\paragraph{Baseline.} We compare \textsc{DUPLEX} against the original \texttt{mini-swe-agent}, \texttt{SWE-agent} \cite{yang2024swe}, \texttt{AutoCodeRover} \cite{zhang2024autocoderover}, and \texttt{OpenHands} \cite{wang2024openhands}. We use three open-source models in our experiments: \texttt{Qwen3-Coder-480B-A35B-Instruct} \cite{qwen3technicalreport}, \texttt{gpt-oss-120b} \cite{agarwal2025gpt}, and \texttt{GLM-4.7-Flash} \cite{5team2025glm45agenticreasoningcoding}. For \texttt{SWE-agent}, we employ the bash-only profile, whereas for all other agents, the default prompt or profile is retained. Each agent is allotted a maximum of 100 steps per task. We use default model parameters across all experiments.

\paragraph{Artifact Use.}
All existing and created artifacts are used or released for research evaluation and reproduction under their applicable public terms.

\subsection{RQ1: Effectiveness}
In this subsection, we present the end-to-end performance of \textsc{DUPLEX} and comparable code-repair agents to discuss the effectiveness of our work.

Table~\ref{tab:rq1_repo_breakdown} summarizes the overall performance of each agent-model combination on the SWE-bench Verified dataset. Overall, \textsc{DUPLEX} achieves the highest end-to-end success rate under all three open-source models considered in this study.

With \texttt{Qwen3-Coder-480B-A35B-Instruct} as the backbone model, \textsc{DUPLEX} reaches a 65.4\% end-to-end task success rate, exceeding \texttt{mini-swe-agent} (55.4\%), \texttt{SWE-agent} (44.2\%), \texttt{AutoCodeRover} (44.2\%), and \texttt{OpenHands} (52.4\%). The margin over the strongest baseline is 10.0 percentage points. At the repository level, it ranks first or tied first on 11 of the 12 repos, with notable results on \texttt{pytest} (84.2\%), \texttt{sklearn} (84.4\%), \texttt{sphinx} (54.5\%), and \texttt{sympy} (66.7\%).

This pattern also holds for the two relatively lightweight open-source models. With \texttt{gpt-oss-120b}, \textsc{DUPLEX} achieves 38.0\%, compared with 26.0\% for \texttt{mini-swe-agent}, 20.8\% for \texttt{SWE-agent}, and 12.6\% for \texttt{AutoCodeRover}; it ranks first or tied first on all 12 repos. With \texttt{GLM-4.7-Flash}, \textsc{DUPLEX} reaches 51.6\%, higher than 46.2\% for \texttt{mini-swe-agent}, 47.8\% for \texttt{SWE-agent}, and 29.4\% for \texttt{AutoCodeRover}; it ranks first or tied first on 9 of the 12 repos, including \texttt{pytest} (73.7\%) and \texttt{sympy} (50.7\%).

Therefore, we conclude that the results indicate that the overall advantage of \textsc{DUPLEX} is consistent across open-source backbones of different scales. We will discuss how \textsc{DUPLEX} achieves this performance in the following subsections. Across the three backbone settings, \textsc{DUPLEX} improves the resolved rate by 3.8-12.0 percentage points over the strongest baseline, with a mean framework-recorded inference cost of \$1.014 per attempted task.

\subsection{RQ2: Patch Compactness}
In software development, overly long pull requests can reduce reviewers’ willingness to perform thorough code reviews. Patch compactness is therefore a relevant practical property, particularly because LLM-generated patches may contain redundant components \cite{abbassi2025taxonomy, santa2025llm}. However, byte size captures only compactness; it does not directly measure correctness, maintainability, security, or human review effort.

We report the mean, median, and 90th percentile of patch sizes, measured in bytes, for successful cases under each agent-model combination. Because different agents may solve different subsets of instances, these conditional statistics can reflect both patch-generation behavior and task mix. We therefore treat them as a descriptive comparison of successful-patch compactness rather than a controlled estimate of patch quality.

\begin{table}[t]
\centering
\footnotesize
\caption{Patch-size statistics for successful repairs, in bytes.}
\label{tab:rq2_patch_size}
\setlength{\tabcolsep}{4.5pt}
\renewcommand{\arraystretch}{1}
\begin{tabular}{llrrr}
\toprule
\textbf{Model} & \textbf{Agent} & \textbf{Mean} & \textbf{Median} & \textbf{P90} \\
\midrule
\multirow{5}{*}{\textbf{Qwen}}
& MSA         & 78,026 & 6,723 & 84,076 \\
& SWA         & 72,033 & 16,753 & 119,421 \\
& ACR         & \textbf{1,283} & \textbf{934} & \textbf{2,532} \\
& OH$^{\dagger}$ & 10,273 & 3,052 & 23,463 \\
& OURS        & 2,231 & 996 & 3,607 \\
\midrule
\multirow{4}{*}{\textbf{GPT}}
& MSA         & 212,466 & 2,261 & 23,778 \\
& SWA         & 220,956 & 2,884 & 21,897 \\
& ACR         & \textbf{1,518} & 1,389 & \textbf{2,331} \\
& OURS        & 53,614 & \textbf{1,127} & 4,494 \\
\midrule
\multirow{4}{*}{\textbf{GLM}}
& MSA         & 5,822 & 1,242 & 13,431 \\
& SWA         & 61,411 & 19,607 & 105,168 \\
& ACR         & \textbf{1,095} & \textbf{814} & \textbf{1,992} \\
& OURS        & 1,374 & 837 & 2560 \\
\bottomrule
\end{tabular}

\raggedright
\footnotesize{$^{\dagger}$ OpenHands is reported only with Qwen; see Table~\ref{tab:rq1_repo_breakdown}.}
\end{table}

Table \ref{tab:rq2_patch_size} shows that the successful patches produced by \textsc{DUPLEX} generally have lower median and 90th-percentile sizes than those produced by \texttt{mini-swe-agent} and \texttt{SWE-agent} across the three backbones. \texttt{AutoCodeRover} remains more compact on most reported statistics, so \textsc{DUPLEX} does not consistently produce the smallest patches. These results should be read alongside the repair rates in RQ1 as two separate properties of the evaluated systems. They show that successful \textsc{DUPLEX} patches are generally compact, but do not establish review readiness or isolate the causal effect of the QA stage.

\subsection{RQ3: Ablation Study}
\begin{figure}[ht]
    \centering
    \includegraphics[width=\columnwidth]{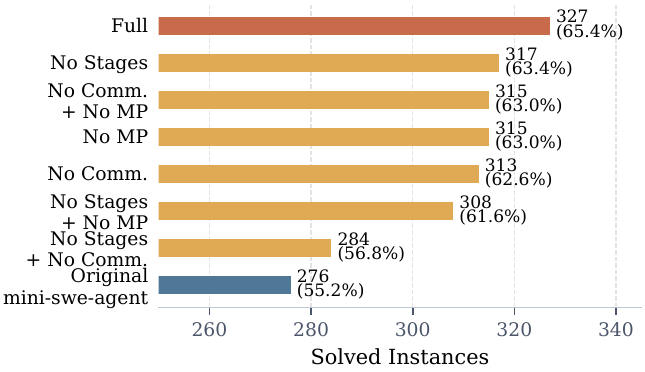}
    \caption{Ablation study of \textsc{DUPLEX} with \texttt{Qwen3-Coder-480B-A35B-Instruct} on SWE-bench Verified. "Comm." and "MP" denote Communication Mechanism, and Monitor Plane, respectively.}
    \label{fig:ablation}
\end{figure}

To understand the contribution of each component in \textsc{DUPLEX}, we conduct an ablation study on SWE-bench Verified using \texttt{Qwen3-Coder-480B-A35B-Instruct}. 
We progressively remove key components from the full system, including the Worker Plane stage decomposition, the Monitor Plane, and the controlled communication mechanism, and report the end-to-end repair performance in Figure~\ref{fig:ablation}.

Figure~\ref{fig:ablation} shows that the full \textsc{DUPLEX} system achieves the strongest performance, solving 327 instances (65.4\%). 
Removing individual components consistently reduces performance. 
Without Worker Plane stages, the score drops to 317 solved instances (63.4\%), indicating that stage decomposition helps preserve phase-specific focus across reproduction, implementation, and quality assurance. 
Removing the Monitor Plane reduces the result to 315 instances (63.0\%), a 2.4-percentage-point decrease under the default configuration ($m=5$, $n=15$).
Removing the communication mechanism leads to 313 solved instances (62.6\%), the largest single-component drop, showing the importance of controlled cross-stage consultation and delegated low-level execution.

The combined ablations further highlight the interaction among the components. 
Removing both the communication mechanism and the Monitor Plane solves 315 instances (63.0\%), while removing the stages and the Monitor Plane solves 308 instances (61.6\%). 
The largest degradation occurs when both stage decomposition and communication are removed, where performance drops to 284 instances (56.8\%), approaching the original \texttt{mini-swe-agent} baseline. 
Overall, these results show that the gain of \textsc{DUPLEX} does not come from a single component, but from the combined effect of staged repair, trajectory monitoring, and controlled communication.

\section{Related Works}\label{s5}
Recent progress in software engineering has led to a broad range of approaches for software repair, ranging from traditional analysis-based methods to recent LLM-driven systems. In this section, we briefly review the lines of research relevant to our work, including traditional software repair, agentic software repair, and agentless repair methods.

\paragraph{Traditional Automatic Software Repair. }
Traditional automatic software repair bases on rule-based transformations, search-based mutation, and program analysis to generate patches \cite{le2011genprog, nguyen2013semfix, mechtaev2016angelix}. However, these methods are usually designed for narrow bug classes or constrained settings, making them less effective for complex repo-level issues described in natural language.

\paragraph{Agentic Software Repair. }
With the rise of LLMs and ReAct-style agent structures \cite{yao2022react}, software repair has increasingly shifted toward interactive agents that interleave reasoning and action on real repositories. Representative works include multi-agent systems such as MAGIS \cite{tao2024magis} and AutoCodeRover \cite{zhang2024autocoderover}, simple ReAct-style agents such as SWE-agent and mini-swe-agent \cite{yang2024swe}, and inference-scaled agents such as DARS \cite{aggarwal2025dars}. Industrial efforts include Alibaba LingmaAgent \cite{ma2025alibaba} and Trae \cite{gao2025trae}. Our work \textsc{DUPLEX} belongs to this category. Related agentic platforms include OpenHands \cite{wang2024openhands} and Refact.ai\footnote{\url{https://refact.ai/}}.

Some works focus on specific stages of issue resolution rather than end-to-end repair. For issue localization, BLAZE \cite{chakraborty2025blaze} uses dynamic code segmentation and Meta-RAG \cite{tawosi2025meta} uses multi-agent retrieval. AEGIS \cite{wang2024aegis} focuses on issue understanding and script generation. To improve repository understanding and source code indexing, some agentic systems adopt graph-based representations that model relations among code elements, such as function calls, inheritance, and control flow. Alibaba LingmaAgent \cite{ma2025alibaba}, Codexgraph \cite{liu2025codexgraph}, and RepoGraph \cite{ouyang2024repograph} all use graphs to support repository understanding at different levels. Some industrial systems, such as Claude Code\footnote{\url{https://claude.ai/}}, do not explicitly organize codes, but rely on command-line tools.

Unlike prior systems that address individual aspects of agentic repair, \textsc{DUPLEX} integrates trajectory-level monitoring, step-aligned context and code restoration, stage-isolated execution, structured artifact handoff, controlled cross-stage communication, and dedicated quality assurance into a unified repair workflow. While Agentless adopts predefined repair stages, MAGIS uses role-based collaboration, AutoCodeRover separates retrieval and patch generation, and OpenHands supports delegated tool execution, \textsc{DUPLEX} connects these mechanisms through a unified control path that coordinates monitoring, repair, and post-repair refinement.

\paragraph{Agentless Methods. }
In contrast to agentic approaches, agentless methods resolve issues without interactive agent execution. For example, Agentless \cite{xia2025demystifying} decomposes issue resolution into a sequence of predefined phases, while SWE-Fixer \cite{xie2025swe} trains lightweight open-source LLMs for software engineering with dedicated file retrieval and code editing modules. To improve source code indexing, existing methods also adopt tree-based representations that summarize repository structure and provide LLMs with layout information. Agentless \cite{xia2025demystifying} uses such a tree-based representation, an approach later adopted by many agentless works.

\section{Conclusion}\label{s6}
In this paper, we introduced \textsc{DUPLEX}, an end-to-end repo-level code repair agent with a two-plane architecture for improved controllability in long-horizon repair. The Monitor Plane provides trajectory-level progress control through stagnation detection and context backoff, while the Worker Plane organizes repair into independent stages with controlled communication. Experiments show that \textsc{DUPLEX} consistently improves end-to-end repair performance across multiple open-source backbone models and outperforms strong baselines by a clear margin. Overall, the results suggest that effective repo-level repair depends not only on model capability, but also on agent architectures that better manage focus, progress, and patch quality.

\section{Limitations}
Although \textsc{DUPLEX} improves repo-level code repair across multiple open-source backbone models, it still has several limitations. First, our evaluation is conducted on SWE-bench Verified, which mainly consists of Python repositories and benchmark-style issue resolutions. Therefore, the results may not fully reflect performance on other programming languages, larger industrial codebases, or tasks that require complex non-code changes. Second, \textsc{DUPLEX} relies on executable feedback, such as reproduction scripts and tests. When the available tests are incomplete or when the issue cannot be reliably reproduced, the agent may still produce patches that appear correct under the benchmark protocol but fail to address the underlying problem. Third, the Monitor Plane and staged Worker Plane introduce additional LLM calls and control overhead, which may increase cost and latency compared with simpler agents.

As \textsc{DUPLEX} executes commands and modifies repositories, practical deployments should use a least-privilege sandbox, such as \texttt{bubblewrap}\footnote{\url{https://github.com/containers/bubblewrap}}, exposing only the task repository and required tools. Separately, because model-generated patches may be incorrect or insecure even when they pass available tests, they should undergo human review before merging.

\bibliography{custom}

\appendix

\end{document}